\documentclass[%
  aip, pop,  
  reprint,%
]{revtex4-1}

\usepackage{physics}
\usepackage{amsmath,amssymb}
\usepackage{amscd}           
\usepackage{mathrsfs}        
\usepackage{eucal}
\usepackage{braket}          
\usepackage{bm}              

\usepackage[x11names]{xcolor}  
\usepackage[pdftex]{graphicx}  

\usepackage[
  pdfusetitle,
  unicode=true,
]{hyperref}
\usepackage{accents}
\usepackage[utf8]{inputenc}
\usepackage[T1]{fontenc}
\usepackage{mathptmx}

\usepackage{comment}   
\usepackage{etoolbox}

\makeatletter
\def\@email#1#2{%
 \endgroup
 \patchcmd{\titleblock@produce}
  {\frontmatter@RRAPformat}
  {\frontmatter@RRAPformat{\produce@RRAP{*#1\href{mailto:#2}{#2}}}\frontmatter@RRAPformat}
  {}{}
}%
\makeatother

\definecolor{BoadillaBlue}{RGB}{71,70,186}

\begin{document}


\title[Clebsch representation of relativistic plasma and generalized enstrophy]{Clebsch representation of relativistic plasma and generalized enstrophy}

\author{Keiichiro Nunotani}
  \email{nunotani.keiichiro20@ae.k.u-tokyo.ac.jp}
\affiliation{%
  Graduate School of Frontier Sciences, University of Tokyo, Kashiwa, Chiba 277-8561, Japan
}%

\author{Zensho Yoshida}
\affiliation{%
  Graduate School of Frontier Sciences, University of Tokyo, Kashiwa, Chiba 277-8561, Japan
}%
\affiliation{%
  National Institute for Fusion Science, Toki, Gifu 509-5292, Japan
}%

\date{\today}

\begin{abstract}
  The theory of relativistic plasmas is attracting interest as a model of high-energy astronomical objects.
  The topological constraints, built in the governing equations, play an essential role in characterizing the structures of plasmas.
  Among various invariants of ideal models, the circulation is one of the most fundamental quantities, being included in other invariants like the helicity.
  The conventional enstrophy, known to be constant in a two-dimensional flow, can be generalized, by invoking Clebsch variables, to the topological charge of a three-dimensional fluid element, which essentially measures circulations.
  Since the relativistic effect imparts space-time coupling into the metric, such invariants must be modified.
  The non-relativistic generalized enstrophy is no longer conserved in a relativistic plasma, implying that the conservation of circulation is violated.
  In this work, we extend the generalized enstrophy to a Lorentz covariant form.
  We formulate the Clebsch representation in relativity using the principle of least action and derive a relativistically modified generalized enstrophy that is conserved in the relativistic model.
\end{abstract}

\maketitle

\section{Introduction}
\label{sec: Introduction}

Kelvin's circulation theorem reveals a hidden invariance (conservation law) beneath the complicated motion of ideal (dissipation free and barotropic) fluids.
The essence of the theorem is to evaluate the circulation along an arbitrary loop \emph{co-moving} with the fluid.
In that respect, the point of the theorem is similar to the mass conservation law where we evaluate the integral of the density over an arbitrary volume element co-moving with the fluid.
In both cases, the essence lies in the duality of the differential form and geometrical object;
the circulation is the pairing of the momentum 1-form and 1-dimensional loop, while the mass is the pairing of the density $n$-form and $n$-dimensional volume element (where $n$ is the dimension of space).
In a charged fluid (plasma),  we replace the momentum by the canonical momentum dressed by the electromagnetic potential; then, the circulation of the canonical momentum is conserved.

There are many cousins of such invariants, and they can be unified as Casimirs (such as helicities, cross helicities, etc.)\,\cite{YoshidaMorrison2014hierarchy,Fukumoto2008crosshelicity}.
The central idea is to combine Lie-dragged quantities with wedge products, and define a Lie-dragged $n$-form  (where $n$ is the dimension of space).  
For example, in an ideal two-dimensional (2D) flow, the vorticity $\omega$ is a Lie-dragged 2-form. 
Dividing $\omega$ by the Lie-dragged 2-form density $\rho$, we obtain a Lie-dragged scalar $\omega/\rho$.  
Then $f(\omega/\rho)\rho$ (where $f$ is an arbitrary scalar function) is a Lie-dragged 2-form;
hence the \emph{cross-enstrophy} $\int_\Omega f(\omega/\rho)\rho $ (where $\Omega$ is the total domain or a co-moving subdomain) is an invariant,
which is a Casimir in the context of Hamiltonian mechanics~\cite{Morrison1998}.
In general, a Casimir $C$ is a member of the center of a Poisson algebra,
i.e., $\{ C, H \}=0$ for every Hamiltonian $H$, implying that $C$ is an intrinsic constant, unchangeable by any kind of energy.

The existence of such a $C$ is only possible in a non-canonical Poisson algebra (or a Hamiltonian system),
i.e., a canonical Poisson algebra only has a trivial center $C$ with a constant value (see Sec.~\ref{sub: Hamiltonian Systems and Casimirs}).
A canonical Poisson algebra can be \emph{reduced} into a non-canonical subalgebra~\cite{MarsdenWeinstein1974reduction}
in which a Casimir represents the gauge symmetry of the reduced variables.
Conversely, some non-canonical system can be inflated to a larger canonical system;
a Casimir $C$ is then \emph{unfrozen} by coupling with a supplemented variable (e.g., $P$)~\cite{YoshidaMorrison2014unfreezing}.
Seen from the canonicalized larger system, the invariance of $C$ is due to the absence of its conjugate variable $P$ in the Hamiltonian; we call such $P$ a \emph{phantom}~\cite{YoshidaMorrison2014hierarchy}.

While ideal (charged) fluid models are non-canonical Hamiltonian systems\,\cite{Morrison1998},
they can be canonicalized by invoking the \emph{Clebsch parametrization} of fluid variables\,
\cite{Serrin1959mathematical,Seliger1968variational, TanehashiYoshida2015gauge, Yoshida2009},
i.e., we can formulate a canonical Hamiltonian system of Clebsch variables that subsumes the non-canonical (charged) fluid system as a subalgebra (see Sec.~\ref{sub: Hamiltonian Systems and Casimirs}).
Interestingly, all Clebsch variables, except one that is the conjugate of the density, are simply Lie-dragged,
so they are convenient for constructing Casimirs\,\cite{YoshidaMorrison2017_springer,YoshidaMorrison2017_PRL}.
The aforementioned cross-enstrophy is defined for 2D fluid, can be generalized to a topological charge that is invariant even in three-dimensional (3D) fluid, and is also a generalization of Kelvin's circulation (see Sec.\ref{sub: Enstrophy for 3D Flow}).

When we consider relativistic ideal fluids, Casimirs may be subject to relativistic modifications.
For example, in the case of helicity, the fact that the conventional helicity is not preserved has been investigated, and the relativistically modified helicity becomes a Casimir~\cite{YoshidaKawazuraYokoyama2014}.
The relativistically modified enstrophy for ordinary 2D fluids is known~\cite{CarrascoFederico2012turbulent,ElingOz2013holographic}, but that for 
3D flows is not.
In this paper, we pursue the relativistic modification of the enstrophy in 3D flows.

The relativistic effects on the vorticity (including magnetic fields) may play an important role, for example, in generating of the seed magnetic field in early cosmology~\cite{KulsrudZweibel2008originB,MahajanYoshida2010seedB}.
Considering the Hamiltonian structure governing the dynamics of an ideal fluid,
the circulation of the canonical momentum, including both vorticity and magnetic fields, cannot emerge from a zero initial value. 
The primary generation mechanism for the magnetic field in the Universe is not given conclusive explanation to date~\cite{KulsrudZweibel2008originB}. 
While existing explanations (e.g., the baroclinic mechanism~\cite{Charney1947baroclinicmechanism} or Biermann battery~\cite{Biermann1950Biermannbattery}) introduce nonideal dynamics~\cite{KulsrudCenOstrikerRyu1997protogalactic,GnedinFerraraZweibel2000generation},
the relativistic effect is shown to generate an effective baroclinic effect~\cite{MahajanYoshida2010seedB}.  

This paper is organized as follows.
In Section \ref{sec: Preliminary}, we briefly review the basic theory of Hamiltonian formalism.
The concepts of epi-2D flow~\cite{YoshidaMorrison2017_springer,YoshidaMorrison2017_PRL} and relativistic helicity~\cite{YoshidaKawazuraYokoyama2014} that inspired this paper are also briefly summarized.
In Section \ref{sec: Relativistic Plasma}, we find the Clebsch representation in relativity.
We need to describe the fluid motion in terms of the Clebsch representation to consider enstrophy in epi-2D fluids.
Then we verify that the conventional enstrophy conservation law is violated in the relativistic plasma in Section \ref{sec: Semi-Relativistic Generalized Enstrophy},
and formulate the relativistically modified enstrophy to be conserved for the relativistic plasma in Section \ref{sec: Relativistic Generalized Enstrophy}.

\section{Preliminary}
\label{sec: Preliminary}

\subsection{Basic Equations and Clebsch Representation}
\label{sub: Basic Equations and Clebsch Representation}

An ideal (dissipation-free) barotropic fluid is described by a particle number density $\rho$ and a velocity field $\boldsymbol{V}$. Their time evolution follows the Euler equation:
\begin{align}
  \partial_t \rho
  &= - \nabla\!\cdot\!(\rho \boldsymbol{V})
  , \label{naive1} \\
  \partial_t \boldsymbol{V}
  &= - (\boldsymbol{V}\!\cdot\!\nabla)\boldsymbol{V} - \frac{1}{m} \nabla \tilde{h}
  , \label{naive2}
\end{align}
with the molar enthalpy $\tilde{h}$,
which is assumed to be a function of only $\rho$ so that we may write $\rho^{-1} \nabla p = \nabla \tilde{h}$ with the pressure $p$.
For simplicity, we assume the 3-torus $\mathbb{T}^3$ as the 3D domain containing this fluid.
Taking $\nabla \times$ of both sides of equation (\ref{naive2}), we obtain the \emph{vorticity equation}
\begin{align}
  \partial_t \boldsymbol{\omega}
  &= \nabla \times (\boldsymbol{V} \times \boldsymbol{\omega})
  \label{vorticity equation}
\end{align}
where $\boldsymbol{\omega} := \nabla\times\boldsymbol{V}$ is the \emph{vorticity} of the fluid. 
Since $\mathrm{rot}\,\mathrm{grad} = 0$, the second term in equation (\ref{naive2}) vanishes in equation (\ref{vorticity equation}).

In this paper, equations (\ref{naive1}) and (\ref{vorticity equation}) are essential.
They claim that $\rho$ and $\boldsymbol{\omega}$ are constant along $\boldsymbol{V}$, respectively.
We investigate the topological constraints that result from these equations.

The physical quantity $C(t)$ defined as
\begin{align}
  C(t) := \int_{\mathbb{T}^3} \boldsymbol{V} \vdot \boldsymbol{\omega} \,\dd^3 x
\end{align}
is called \emph{helicity}.
This corresponds to the entanglement (the Gauss linking number) of the vortex filaments~\cite{Moffatt1978field,Moffatt1992helicity,YoshidaKawazuraYokoyama2014},
and is conserved for the ideal barotropic fluid.

By introducing virtual potential fields $\varphi,\lambda^1,\sigma_1,\lambda^2,\sigma_2$,
the velocity field $\boldsymbol{V}$ is represented as
\begin{align}
  m \boldsymbol{V} &:= \nabla\varphi + \lambda^1 \nabla\sigma_1 + \lambda^2 \nabla\sigma_2
  \label{conventional P}
\end{align}
with the equation of motion
\begin{subequations}
  \label{intro-Clebsch}
  \begin{align}
    \partial_t \rho + \nabla\cdot( \rho \boldsymbol{V} ) &= 0
    \label{intro-Clebsch rho}
    , \\
    (\partial_t + \boldsymbol{V}\cdot\nabla) \, \lambda^k &= 0
    \label{intro-Clebsch lambda}
    , \\
    (\partial_t + \boldsymbol{V}\cdot\nabla) \, \sigma_k &= 0
    \label{intro-Clebsch sigma}
    , \\
    (\partial_t + \boldsymbol{V}\cdot\nabla) \, \varphi &= \frac{1}{2} m V^2 - \tilde{h}
    \label{intro-Clebsch varphi}
    . 
  \end{align}
\end{subequations}
This expression is called the \emph{Clebsch representation}, and the virtual potential fields above are called the \emph{Clebsch parameters} or the \emph{Clebsch variables}. 
Mathematically, a general 3-dimensional vector field can be cast into the form of equation (\ref{conventional P}); see Ref.~\onlinecite{Yoshida2009}.
Equation (\ref{naive1}) is equivalent to equation (\ref{intro-Clebsch rho}), and equation (\ref{naive2}) is derived from equations (\ref{intro-Clebsch lambda}) -- (\ref{intro-Clebsch varphi}).
Note that the left-hand sides of the equations (\ref{intro-Clebsch}) are the change along $\boldsymbol{V}$ of each parameter.

We will see how Clebsch expressions can be useful in the next subsection.

\subsection{Hamiltonian Systems and Casimirs}
\label{sub: Hamiltonian Systems and Casimirs}

A \emph{Hamiltonian system} is a dynamical system whose time evolution in a phase space is described by
\begin{align}
  \dv{t} \bm{z} = J_{\boldsymbol{z}} \, \partial_z H(\boldsymbol{z})
\end{align}
where $ \boldsymbol{z}$ is the coordinate of the phase space,
the \emph{Hamiltonian} $H$ is a function of $\boldsymbol{z}$, 
and the \emph{Poisson operator} $J_{\boldsymbol{z}}$ is an anticommutative operator with the Jacobi identity.
In particular, the Hamiltonian system is said to be \emph{canonical} when $J_{\boldsymbol{z}}$ is
\begin{align}
  J_{\boldsymbol{z}} = \mqty( 0 & I \\ -I & 0)
\end{align}
where $I$ is an identity operator.

We may consider a more general system (often called a non-canonical Hamiltonian system) such that $\boldsymbol{z}$ has $K$ components and $\mathrm{rank}( J_{\boldsymbol{z}} ) = 2N \leq K$.
From a generalization of the Darboux theorem proven by Lie~\cite{Eisenhart1961continuous,Littlejohn1982mathematical}, $J_{\boldsymbol{z}}$ can be transformed (locally) into the following form:
\begin{align}
  J_{\boldsymbol{z}}
  =
  \mqty(
    0_N   & I_N   & 0 \\ 
    -I_N  & 0_N   & 0 \\
    0     & 0     & 0_{K-2N}
  )
  .
\end{align}
Corresponding to the existence of $K-2N$ degenerate variables, there are an equal number of invariants, called \emph{Casimirs}~\cite{Morrison1998}.
In the same way, a Casimir is defined for the non-canonical Hamiltonian system with infinite degrees of freedom.

The Hamiltonian system described by $(\rho, \boldsymbol{V})$ is known to be non-canonical and the total number of particles and the helicity together are a Casimir~\cite{Morrison1998}.
However, this non-canonical Hamiltonian system is a sub-system of the canonical Hamiltonian system defined below~\cite{YoshidaMorrison2017_springer}.

By introducing Clebsch parameters, the phase space is defined as
\begin{align}
  \Set{
    \xi = \qty( \rho,\varphi,\Lambda^1,\sigma_1,\Lambda^2,\sigma_2 )^\mathrm{T}
  }
  \label{phase-space}
\end{align}
with $\Lambda^k := \rho \lambda^k$,
and the Hamiltonian is defined as
\begin{align}
  H(\xi)
  &:=
  \int_{\mathbb{R}^3} \qty[
    \, \frac{1}{2} mV^2 + \varepsilon(\rho)
  ] \rho\,\dd x^3
\end{align}
where the molar internal energy $\varepsilon$ is associated with the molar enthalpy $\tilde{h}$ by the relation $\tilde{h}(\rho) = \pdv{(\rho \varepsilon(\rho))}{\rho}$.
Substituting these into the Hamilton equation
\begin{gather}
  \dv{t} \xi(t) = J \,\partial_\xi H \,|_{\xi(t)}
  , \\
  J := J_c \oplus J_c \oplus J_c ,\quad J_c := \mqty(0 & I \\ -I & 0)
  ,
\end{gather}
the equations (\ref{conventional P}) and (\ref{intro-Clebsch}) can be derived. 
Therefore, the non-canonical Hamiltonian system defined by the Euler equation is a subsystem of the canonical Hamiltonian system defined by the Clebsch representation.

Here we note that the conserved quantities seen as Casimir in $(\rho, \boldsymbol{V})$ formulation are reinterpreted as the \emph{gauge symmetry} in the Clebsch representation.
See Ref.~\onlinecite[Sec. 3.2]{YoshidaMorrison2017_springer} for a detailed discussion.

\subsection{Plasmas Modeled as Charged Fluids}
\label{sub: Plasmas Modeled as Charged Fluids}

When the fluid is electrically charged, its time evolution is written as
\begin{align}
  \partial_t \rho + \nabla\!\cdot\!(\rho \boldsymbol{V})
  &= 0
  , \label{naive3} \\
  (\partial_t + \boldsymbol{V}\!\cdot\!\nabla) P_i
  &= - \partial_i \qty( \tilde{h} + \frac{e}{c} \phi ) + \frac{e}{c} V_j \partial_i A_j
  , \label{naive4}
\end{align}
where $\boldsymbol{P} := m \boldsymbol{V} + (e/c) \boldsymbol{A}$ is the canonical momentum field
with the mass $m$ and the charge $e$. 
The second term on the right-hand side of equation (\ref{naive4}) is unfamiliar, but it can be rewritten as follows:
\begin{align}
  \partial_t P_i + (\boldsymbol{V}\!\cdot\!\nabla) P_i
  - V_j \partial_i P_j
  &= - \partial_i \qty( \frac{1}{2}mV^2 + \tilde{h} + \frac{e}{c} \phi )
  \label{naive5}
  . 
\end{align}
The left-hand side of equation (\ref{naive5}) expresses how much $\boldsymbol{P}$ varies along $\boldsymbol{V}$, corresponding to the Lie derivative $\mathcal{L}_{\boldsymbol{V}} \boldsymbol{P}$ of the 1-form $\boldsymbol{P}$ (see Appendix~\ref{sec: Notation Comparison of Differential Geometry and Tensor Analysis}).
By redefining $\boldsymbol{\omega} := \nabla \times \boldsymbol{P}$, the equation (\ref{vorticity equation}) remains valid.
By changing the left-hand side of equation (\ref{conventional P}) to $\boldsymbol{P}$ and adding $-\phi$ to the right-hand side of equation (\ref{intro-Clebsch varphi}), the charged fluid can also be represented by Clebch parameters.
The electromagnetic potentials $\phi, \boldsymbol{A}$ obey Maxwell equations with appropriate charges and currents.

In this paper, the term ``plasma'' refers to a system of charged fluids, i.e., a mixture of single or multiple electrically charged ideal fluids.
We especially focus on a single charged fluid and discuss the pertinent conservation laws. 

For a plasma consisting of multiple species of charged fluids (e.g., electrons and protons), a set of Clebsch parameters is required for each fluid species.
The Clebsch parameters belonging to each set are still Lie dragged, while the number of four-current terms in Maxwell equations increases with the number of fluid species.
Since Maxwell equations are not used to derive the enstrophy conservation law, even for multiple species model, enstrophy are conserved for each fluid species.
In other words, the enstrophy conserves no matter how the charged fluid of interest is moved by the electromagnetic field representing the interactions among charged fluids.

We note that the vector field that transports each fluid element becomes an issue in the actual application of the conservation laws.
The limit to the extended magnetohydrodynamics (exMHD), Hall MHD, or MHD, must take into account the charge neutrality condition to avoid singularity.
In MHD, the magnetic field is frozen into the fluid (Alfv\'{e}n's theorem)~\cite{Alfven1942frozen}.
The circulation conservation law for the canonical momentum field in this study corresponds to Alfv\'{e}n's theorem in MHD, because the inertial term of the electron is neglected among the two fluids in MHD.
When interpreting the meaning of frozen magnetic field lines in relativistic model, the time resetting projection plays an important role~\cite{Pegoraro2012relativisticfrozen,ComissoAsenjo2020vorticity}.

As mentioned in Sec. \ref{sub: Basic Equations and Clebsch Representation}, it is essential that $n = \rho\,\dd^3 x$ (3-form) and $\omega$ (2-form) are Lie-dragged, i.e., conserved against the action of the vector $\boldsymbol{V}$. In the ``real world,’’ there are many mechanisms that break such conservation laws.  For example, in the presence of pair production or annihilation, the particle number conservation is broken, and $\rho$ is no longer conserved. A finite collisionality or the radiation reaction effect brings about non-exact terms to the right-hand side of the momentum equation (\ref{naive5}), and dissipation terms appear in the vorticity equation  (\ref{vorticity equation}). Understanding the fundamental conservation laws and their topological implication helps us to see how such ``non-ideal effects’’ work to free the topological constraints and enable the universe to evolve. 

The behavior of a neutral ideal fluid and a plasma modeled as charged fluids can be discussed in parallel.
In the following, we will describe the plasma.

\subsection{Enstrophy for 3D Flow}
\label{sub: Enstrophy for 3D Flow}

For a 2D flow, helicity is clearly identically zero.
For example, if there is no flow in the $z$-direction and the physical quantity has no gradient in the $z$-direction, then
\begin{align}
  \boldsymbol{V} &= (v_x, v_y, 0)^\mathrm{T}
  , \\
  \omega &= \nabla\times m \boldsymbol{V} = m (\partial_x v_y - \partial_y v_x) \boldsymbol{e}_z
  \label{2D_omega}
  .
\end{align}
Since $\boldsymbol{V}$ and $\boldsymbol{\omega}$ are orthogonal, the helicity $C$ is identically zero.
Instead of helicity becoming meaningless, another constant of motion, the \emph{generalized enstrophy}, appears.
It is defined as
\begin{align}
  \int_{\Sigma(t)} f \qty( \frac{\omega_z}{\rho} ) \, \rho \dd^2 x
  \label{2D_generalized_enstrophy}
\end{align}
where $\omega_z = m (\partial_x v_y - \partial_y v_x)$ considering equation (\ref{2D_omega}), $f : \mathbb{R} \to \mathbb{R}$ is an arbitrary smooth function, and $\Sigma(t) \subset \mathbb{R}^2$ is an arbitrary region that co-moves with the (non-relativistic) 2D fluid.
For the simple choice $f = \mathrm{id}_{\mathbb{R}}$, equation (\ref{2D_generalized_enstrophy}) is rewritten as 
\begin{align}
  \int_{\Sigma(t)} \omega_z \, \dd^2 x
  = \int_{\partial\Sigma(t)} m \boldsymbol{V} \cdot \dd \boldsymbol{x}
  .
  \label{2D_generalized_enstrophy_id}
\end{align}
The left-hand side of equation (\ref{2D_generalized_enstrophy_id}) represents the number of the vortex filaments through $\Sigma(t)$,
and the constancy of the right-hand side is known as Kelvin's circulation theorem.
For a density constant flow, equation (\ref{2D_generalized_enstrophy}) has a special form of $\int_{\Sigma(t)} (\omega_z)^2 \, \dd^2 x$ conventionally called \emph{enstrophy}.
This is why the term ``generalized enstrophy'' includes the adjective ``generalized.''

To extend the concept of generalized enstrophy to general 3D flow, it is necessary to consider the vortex decomposed into two parts according to the Clebsch representation.
For a 3D flow $\boldsymbol{V}$ represented by equation (\ref{conventional P}), we define $\widetilde{Q}_1(t)$ and $\widetilde{Q}_2(t)$ as
\begin{subequations}
  \begin{align}
    \widetilde{Q}_1(t) &:= \int_{\Omega(t)} f(\vartheta_1) \, \rho \, \dd^3 x
    , &
    \vartheta_1 := \frac{\, \boldsymbol{\omega}_1 \cdot \nabla\sigma_2 \,}{\rho}
    \\
    \widetilde{Q}_2(t) &:= \int_{\Omega(t)} f(\vartheta_2) \, \rho \, \dd^3 x
    , &
    \vartheta_2 := \frac{\, \boldsymbol{\omega}_2 \cdot \nabla\sigma_1 \,}{\rho}
  \end{align}
\end{subequations}
where $\boldsymbol{\omega}_k := \nabla\lambda^k \!\times\nabla\sigma_k$ is the component of the decomposition of vorticity $\boldsymbol{\omega}$,
$f : \mathbb{R} \to \mathbb{R}$ is an arbitrary smooth function again,
and $\Omega(t) \subset \mathbb{R}^3$ is an arbitrary region that co-moves with the (non-relativistic) fluid.
Both $\widetilde{Q}_1(t)$ and $\widetilde{Q}_2(t)$ conserve because they are the product of the Lie-dragged Clebsch parameters, i.e.,
\begin{subequations} \label{conventional_enstrophy}
  \begin{align}
    \widetilde{Q}_1(t) := \int_{\Omega(t)} f(\vartheta_1) \, n
    ,\,\,
    \vartheta_1 := \frac{ \qty( \omega_1 \wedge \dd\sigma_2 )^\ast }{n^\ast}
    ,\,\,
    \omega_1 := \dd\lambda^1 \!\wedge \dd\sigma_1
    \\
    \widetilde{Q}_2(t) := \int_{\Omega(t)} f(\vartheta_2) \, n
    ,\,\,
    \vartheta_2 := \frac{ \qty( \omega_2 \wedge \dd\sigma_1 )^\ast }{n^\ast}
    ,\,\,
    \omega_2 := \dd\lambda^2 \!\wedge \dd\sigma_2
  \end{align}
\end{subequations}
with a 3-form $n := \rho \, \dd x^3$
in differential geometry notation (see Appendix~\ref{sec: Notation Comparison of Differential Geometry and Tensor Analysis}).

As a counterpart to a 2D flow, suppose a fluid that is uniform in the $z$ direction and moving only in the $xy$ directions in 3D space.
If we take $\sigma_2 = z$, then $\widetilde{Q}_1(t)$ coincides with the generalized enstrophy of the 2D fluid~\cite{YoshidaMorrison2017_springer}.
In this sense, $\widetilde{Q}_k(t)$ defined here is an extension of the enstrophy in a 2D flow.
Even if $\boldsymbol{V}$ is not a 2D flow embedded in 3D space, in a region where either $\lambda^1$ or $\lambda^2$ is zero, $\widetilde{Q}_2(t)$ or $\widetilde{Q}_1(t)$ represents the number of vortex filaments as well as the conventional generalized enstrophy.
However, in regions where neither $\lambda^1$ nor $\lambda^2$ is zero, $\widetilde{Q}_k(t)$ does not represent an observable quantity because the Clebsch representation is not unique for a given flow.

The idea of extending the enstrophy to a 3D flow comes from Yoshida and Morrison~\cite{YoshidaMorrison2017_springer,YoshidaMorrison2017_PRL}.
In the paper, the intermediate class between 2D and 3D flows, termed \emph{epi-two-dimensional flow} (\emph{epi-2D flow}), was introduced,
and the helicity-enstrophy interplay was investigated.

Since the arguments for $\widetilde{Q}_1(t)$ and $\widetilde{Q}_2(t)$ are parallel, only $\widetilde{Q}_1(t)$ will be discussed below;
for this reason, $\widetilde{Q}_1(t)$ will be abbreviated as $\widetilde{Q}(t)$.

\subsection{Relativistic Helicity}
\label{sub: Relativistic Helicity}

Whereas so far we have looked at helicity in non-relativistic fluids, a relativistic helicity has been formulated by Yoshida, Kawazura, and Yokoyama~\cite{YoshidaKawazuraYokoyama2014}.
They showed that the conventional helicity is no longer conserved by the relativistic effect, and formulated the relativistically modified helicity that is conserved in the relativistic model.
In this subsection, we use differential geometry notation, following Ref.~\onlinecite{YoshidaKawazuraYokoyama2014}
(see Appendix~\ref{sec: Notation Comparison of Differential Geometry and Tensor Analysis} for differential geometry notation in detail).

Let $\mathcal{U} := \gamma \, ( c \, \partial_0 + V^i \partial_i )$ be a proper velocity with the Lorentz factor $\gamma$, and $A^\mu$ be the electromagnetic (EM) potential.
While the 1-form $\mathcal{P} := [ (h/c^2) \,\mathcal{U}_\mu + (e/c) A_\mu ] \, \dd x^\mu$ represents the canonical momentum field, the 2-form $\dd\mathcal{P}$ is the relativistic counterpart of the vorticity.
The components of $\dd\mathcal{P}$ are
\begin{align}
  \partial^\mu \mathcal{P}^\nu - \partial^\nu \mathcal{P}^\mu
  = \frac{1}{c^2} \qty[
    \partial^\mu \qty( h\,\mathcal{U}^\nu ) - \partial^\nu \qty( h\,\mathcal{U}^\mu )
  ]
  + \frac{e}{c} F^{\mu\nu}
\end{align}
with the EM tensor $F_{\mu\nu} := \partial_\mu A_\nu - \partial_\nu A_\mu$%
\footnote{%
  The relativistic vorticity 2-form $\dd\mathcal{P}$ is similar to the generalized magnetofluid field tensor $\mathcal{M}^{\mu\nu}$ presented in Comisso and Asenjo~\cite{ComissoAsenjo2020vorticity}. 
  Please compare with equation (30) in Ref.~\onlinecite{ComissoAsenjo2020vorticity}. 
  Note that Comisso and Asenjo dealt with MHD, which is different from the charged fluid model in this paper.
}. 
Let us denote the component $\mathcal{K}^\mu$ of 3-form $\mathcal{K} := \mathcal{P} \wedge \dd\mathcal{P}$.
The non-relativistic helicity can be expressed as follows:
\begin{align}
  C(t)
  = \int_{X(t)} \mathcal{P} \wedge \dd\mathcal{P}
  = \int_{X(t)} \mathcal{K}^0 \,\dd^3 x
\end{align}
This integral domain $X(t)$, which is a 3-dimensional plane ($t$-plane) in a 4-dimensional Lorentzian manifold, is not Lorentz invariant.
Its time development is
\begin{align}
  \dv{t} C(t) = -2 \int_{X(t)} \theta \mathcal{B} \vdot \nabla \gamma^{-1} \,\dd^3 x
\end{align}
where $\mathcal{B} := \nabla\times (\mathcal{P}^i)$ denotes the spatial components of $\dd\mathcal{P}$~\cite{YoshidaKawazuraYokoyama2014}.
We also consider quasi-static thermal interactions that can be described only by a scalar function $\theta$, which satisfies $T \dd S = \dd\theta$ with the entropy $S$ and temperature $T$.
$C(t)$ is conserved in the non-relativistic limit $\gamma\to1$, but not in general.

Then, the relativistic helicity is defined as follows:
\begin{align}
  \mathfrak{C}(s) := \int_{V(s)} \mathcal{P} \wedge \dd\mathcal{P}
  .
\end{align}
The integral domain $V(s)$ is defined as the 3-dimensional manifold where the domain $V_0$ on the $t$-plane is pushed away by the vector field $\mathcal{U}$ for a proper time $s$, so it is Lorentz invariant.
Indeed,
\begin{align}
  \dv{s} \mathfrak{C}(s)
  = c^{-1} \int_{\partial V(s)}
    \qty(
      h + \frac{e}{c} \, i_\mathcal{U} A - \theta
    )
    \dd\mathcal{P}
  ,
\end{align}
and if $\mathrm{supp}\,\dd\mathcal{P} \cap X(0) \subset V_0$ , then $\mathfrak{C}(s)$ is a constant of motion~\cite{YoshidaKawazuraYokoyama2014}.

\section{Relativistic Plasma}
\label{sec: Relativistic Plasma}

In preparation for considering the enstrophy, in this section we will derive the Clebsch representation in relativistic plasmas.
Since the Clebsch representation in non-relativistic theory is naturally derived from the principle of least action~\cite{Serrin1959mathematical,Seliger1968variational,YoshidaMahajan2011}, we will obtain the relativistic Clebsch representation by defining the Lorentz invariant action appropriately and then computing the principle of least action.
In calculating the variational principle, it is easier to write each physical quantity as components rather than in differential forms, so we use the notation of tensor calculations in this section.

Let $M = \mathbb{R}^4$ be the Minkowski space of signature $(+,-,-,-)$.
We introduce a natural coordinate $x^\mu = (ct,x,y,z)$ such that the metric tensor $\eta$ is represented by
$\eta^{\mu\nu} = \eta_{\mu\nu} = \mathrm{diag}\{+1,-1,-1,-1\}$, where $c$ is the speed of light.
In this paper, we consider plasma which consists of a single kind of charged particles, whose rest mass is $m$ and whose charge is $e$.\footnote{
  This charge $e$ is NOT necessarily the elementary charge $1.602 \times10^{-19} ~\mathrm{C}$.
}

The plasma behavior is represented by the number density field $n(x)$, the velocity field $v^j(x)$, and the EM potential $A^\mu(x)$.
The non-relativistic four-velocity field $u^\mu(x)$ and the relativistic four-velocity field $\mathcal{U}^\mu(x)$ are defined as
\begin{align}
  u^\mu(x) &:= (c, v^j(x))
  , &
  \mathcal{U}^\mu(x) &:= \gamma(x) u^\mu(x)
  \label{mathcalU} ,
\end{align}
with the Lorentz factor
\begin{align}
  \gamma(x) := \qty( 1 - \frac{\boldsymbol{V}(x)^2}{c^2} )^{-1/2}
  \label{gamma} .
\end{align}
Here, let us pay attention to $\norm{\,\mathcal{U}^\mu(x)}_{\eta} \equiv c$, denoting the norm induced from $\eta$ by $\norm{\cdot}_{\eta}$.

The molar internal energy $\mathcal{E}$ includes the rest mass energy $mc^2$ and the thermal energy~\cite{LandauLifshitz1987fluid}.
In this chapter, we assume that $\mathcal{E}$ depends only on the number density $n$ at each point, thus we denote it as $\mathcal{E}(n)$.
The molar enthalpy $h(n)$ is defined by
\begin{align}
  h
  := \pdv{(n\mathcal{E}(n))}{n}
  = \mathcal{E} + \frac{1}{n} \, p
  \label{enthalpy}
\end{align}
where $p$ is the pressure.
The relationship with $p$ comes from the thermodynamic equation $\mathcal{E} = \int p \, \dd\qty( n^{-1} )$.

Here we introduce Clebsch parameters $\varphi(x)$, $\lambda^k(x)$, $\sigma_k(x)$ $(k=1,2)$ for relativistic plasma, imitating the method for non-relativistic plasma \cite{YoshidaMahajan2011}.
These scalar functions place constraints on trajectories in the plasma's phase space in the form of Lagrange multipliers.

The Lagrangian density of non-relativistic plasma is known as
\begin{align}
  \widetilde{L}_F
  &=
  \qty[
    \frac{1}{2}mV^2 - \varepsilon
    - \frac{e}{c}(\phi - \boldsymbol{V} \!\vdot\! \boldsymbol{A})
    - D_t \varphi - \sum_k \lambda^k D_t \sigma_k
  ] n
  \label{non-rel L_F}
  , \\
  \widetilde{L}_{EM}
  &=
  \frac{1}{2c} E^2 - \frac{1}{2} B^2
  \label{non-rel L_EM}
\end{align}
where $D_t$ is the Lagrange derivative $D_t := \partial_t + \boldsymbol{V}(x) \!\vdot\! \nabla = \gamma(x)^{-1} \mathcal{U}^\mu(x) \partial_\mu$.
By analogy to the non-relativistic case, we will define the Lagrangian density as follows:
\begin{align}
  L_F &
  :=
  - n \, \mathcal{U}^\mu
    \qty(
      \frac{1}{c^2} \mathcal{E}(n) \, \mathcal{U}_\mu + \frac{e}{c} A_\mu
      + \partial_\mu \varphi
      + \sum_k \lambda^k \partial_\mu \sigma_k
    )
  \notag \\ &
  =
  - n \qty[
    \mathcal{E}(n) + \mathcal{U}^\mu
    \qty(
      \frac{e}{c} A_\mu
      + \partial_\mu \varphi
      + \sum_k \lambda^k \partial_\mu \sigma_k
    )
  ]
  \label{L_F}
  , \\
  L_{EM} &:= - \frac{1}{4} F^{\mu\nu} F_{\mu\nu}
  \label{L_EM}
\end{align}
with the EM tensor $F_{\mu\nu} := \partial_\mu A_\nu - \partial_\nu A_\mu$.
Then the action functional of the relativistic plasma is defined as
\begin{align}
  S\qty[n,v^j,\varphi,\lambda^k,\sigma_k,A^\mu]
          := \int (L_F(x) + L_{EM}(x)) \, \dd^4 x
  .
  \label{action}
\end{align}
Notice that arguments of this functional are not $\mathcal{U}^\mu(x)$ but $v^j(x)$.
The possible values of $\mathcal{U}^\mu(x)$ have to meet two conditions; the 0th value $\mathcal{U}^0(x)$ should be positive, and the Lorentz norm of $\mathcal{U}^\mu(x)$ should be $c$.
It is difficult to calculate variation of the functional under such complicated conditions.
Conversely, $v^j(x)$ is a parametrization of $\mathcal{U}^\mu(x)$, and is free to assume values in the \emph{open} 3-ball of radius $c$.
Thus, we choose $v^j(x)$ as arguments of the action functional.

Applying the principle of least action to the action functional defined above, we obtain the following results:
\begin{subequations}
\begin{align}
  \partial_\mu (n\,\mathcal{U}^\mu)       &= 0
  \label{delta varphi} , \\
  \mathcal{U}^\mu \partial_\mu \lambda^k  &= \mathcal{U}^\mu \partial_\mu \sigma_k = 0
  \label{delta lambda sigma} , \\
  \mathcal{U}^\mu \partial_\mu \varphi    &= - h(n) - \frac{e}{c} \,\mathcal{U}^\mu A_\mu
  \label{delta n} , \\
  \Box A_\mu - \partial_\mu(\partial_\nu A^\nu) &= \frac{e}{c} n\,\mathcal{U}_\mu
  \label{delta A} , \\
  \frac{1}{c^2} h(n) \,\mathcal{U}_\mu + \frac{e}{c} A_\mu &+ \partial_\mu \varphi + \sum_k \lambda^k \partial_\mu \sigma_k = 0
  \label{delta v} .
\end{align}
\end{subequations}
See Appendix~\ref{sub: Variation of the Action Functional to Derive Clebsch Parametrization} for intermediate calculations.
Equations (\ref{delta varphi}) and (\ref{delta A}) represent the particle number conservation and Maxwell equations, respectively.
Equations (\ref{delta lambda sigma}) and (\ref{delta n}) show the time evolution of Clebsch parameters.
Equation (\ref{delta v}) shows the correspondence between the Clebsch parameters and real physical quantities.

Based on these equations, the canonical momentum field of the relativistic plasma is defined as follows:
\begin{align}
  \mathcal{P}^\mu := \frac{1}{c^2} h(n) \,\mathcal{U}^\mu + \frac{e}{c} A^\mu
  \label{P-canonical} .
\end{align}
With this definition, the canonical momentum can be written in terms of the scalar fields $\varphi(x)$, $\lambda^k(x)$, and $\sigma_k(x)$, according to equation (\ref{delta v}):
\begin{align}
  \mathcal{P}_\mu = - \partial_\mu \varphi - \sum_k \lambda^k \partial_\mu \sigma_k
  \label{P-canonical Clebsch} .
\end{align}
Calculating the Lagrange derivative of the canonical momentum gives
\begin{align}
  \mathcal{U}^\nu (\partial_\mu \mathcal{P}_\nu - \partial_\nu \mathcal{P}_\mu) = 0
  .
\end{align}
This corresponds to the equation of motion $i_\mathcal{U}\dd\mathcal{P} = 0$ as in Ref.~\onlinecite{YoshidaKawazuraYokoyama2014}.

Using equations (\ref{delta varphi})--(\ref{delta n}), (\ref{delta v}) and the definition of the enthalpy $h$, we can derive the generalized Ohm's law
\begin{align}
  \partial_\nu \qty[
    \, \frac{1}{c^2} n h(n) \, \mathcal{U}^\mu \mathcal{U}^\nu 
  ]
  &
  = \frac{e}{c} n \,\mathcal{U}_\nu F^{\mu\nu}
    + \partial^\mu p
  \label{generalized_Ohm} . 
\end{align}
The intermediate calculations are described in Appendix~\ref{sub: Derive the Generalized Ohm's Law from Clebsch Parametrization}.
In this formulation, $(1/c^2) n h(n) \, \mathcal{U}^\mu \mathcal{U}^\nu$ corresponds to the energy-momentum tensor of fluid.
Equation (\ref{generalized_Ohm}) is an expression for the energy-momentum balance when the interaction with the EM field is regarded as an external force.

Using the canonical momentum $\mathcal{P}$ defined earlier, equation (\ref{generalized_Ohm}) can also be rewritten as follows:
\begin{align}
  n \, \mathcal{U}^\nu \partial_\nu \mathcal{P}^\mu
  = \frac{e}{c} n \, \mathcal{U}_\nu \partial^\mu A^\nu + \partial^\mu p
  \label{relativistic_Euler}
  .
\end{align}
For neutral ideal fluids ($e=0$), this corresponds to the Euler equation.

In equation (\ref{P-canonical Clebsch}), the sign of the Clebsch representation is reversed from that of equation (\ref{conventional P}) in the non-relativistic case.
This is because the three components of the non-relativistic 1-form correspond to the spatial components of the 1-form in relativistic theory, and the sign of the spatial components of the Lorentz metric $\eta$ is negative.
Except for the difference in sign style, the Clebsch representation derived here is consistent with previous studies such as Ref.~\onlinecite{KawazuraMiloshevichMorrison2017}.\footnote{
  The model in this study is relativistic charged fluid, whereas the model in Ref.~\onlinecite[Sec. II]{KawazuraMiloshevichMorrison2017} is relativistic extended magnetohydrodynamics.
}

Finally, we confirm that the relativistic generalized Ohm's law can be derived from the Clebsch representation obtained in this section.

\section{Semi-Relativistic Generalized Enstrophy}
\label{sec: Semi-Relativistic Generalized Enstrophy}

Using the Clebsch representation obtained in the previous section, we investigate how generalized enstrophy behaves in the framework of relativity.
The first step is to rewrite the non-relativistic generalized enstrophy for the framework of relativism.

In the previous section, we used the principle of least action as our starting point, but in this section, we will use the Clebsch representation derived in the previous section with the term for thermodynamic effects added. Please note that there is a leap of logic here.

In the following, we will use a formulation based on differential geometry because it offers a number of benefits, including the ability to distinguish between the Lie derivative as a 0-form and the Lie derivative as a 4-form.
The meaning of the symbols used in this paper is defined in Appendix~\ref{sec: Notation Comparison of Differential Geometry and Tensor Analysis}.

Let $M$ be a general $C^\infty$ manifold of dimension $n$ with a volume form $\mathrm{vol}^n$.
For $\alpha \in \Omega^n(M)$, $\alpha^\ast \in \Omega^0(M) = C^\infty(M)$ uniquely exists such that $\alpha = \alpha^\ast \mathrm{vol}^n$.
This notation comes from the Hodge star, which is an isomorphic mapping from $k$-forms to $(n-k)$-forms.

Now that we have defined the mathematical terms, let us consider the physical setting.
The space-time manifold $M = \mathbb{R}^4$ with a global coordinate $(x^0, x^1, x^2, x^3)$, the metric tensor $\eta$, light speed $c$, and charge $e$ are assumed to be the same as in the previous section.
We introduce a volume form $\mathrm{vol}^4 := \dd x^0 \!\wedge\! \dd x^1 \!\wedge\! \dd x^2 \!\wedge\! \dd x^3 \in \Omega^4(M)$ on $M = \mathbb{R}^4$.

Let $\Set{ \qty( n, \varphi, \lambda^1, \sigma_1, \lambda^2, \sigma_2, A ) }$ be a phase space where $n \in \Omega^4(M)$, $\varphi, \lambda^k, \sigma_k \in \Omega^0(M)$, and $A \in \Omega^1(M)$.
$n$ is closely related to the number density 
(more precisely, $i_\mathcal{U} n \leftrightarrow \rho \mathcal{U}^\mu$ is the number density \emph{flux} when $n = \rho\,\mathrm{vol}^4$, and its 0th component is the number density) 
and $A^\mu$ is interpreted as the electromagnetic potential.
The others are Clebsch parameters, which are constraints in the phase space.
Again, we assume that the molar internal energy $\mathcal{E}$ depends only on $n$, and define the molar enthalpy $h$ by equation (\ref{enthalpy}).
The canonical momentum field $\mathcal{P}$ is defined as $\mathcal{P} := - \dd\varphi - \sum_k \lambda^k \dd\sigma_k \in \Omega^1(M)$ as follows from equation (\ref{P-canonical Clebsch}).
The relativistic four-velocity field $\mathcal{U}^\mu \in \mathfrak{X}(M)$ is defined as the dual of the 1-form $U := \frac{c^2}{h} \qty( \mathcal{P} - \frac{e}{c} A ) \in \Omega^1(M)$, i.e., the relation $U = -\!\ast\!(i_\mathcal{U} \mathrm{vol}^4)$ holds where $\ast$ denotes a Hodge star.
Concerning the components, we have $\mathcal{U} = \mathcal{U}^\mu \partial_\mu$ while $U = \mathcal{U}_\mu \dd x^\mu$.
The non-relativistic four-velocity field $u$ in the global coordinate $(x^0, x^1, x^2, x^3)$ is defined as $u := ( c / \,\mathcal{U}^0 ) \,\mathcal{U} \in \mathfrak{X}(M)$.
Of course, $\mathcal{U}^0$ depends on the coordinate system, but here the coordinate system is fixed at $(x^0, x^1, x^2, x^3)$.

In the following, we assume that the plasma is a barotropic fluid, that is, the entropy $S$ is a function of the temperature $T$.
In this case, there exists a scalar function $\theta$ such that $T \dd S = \dd \theta$.
The following differential equations represent the time evolution in the phase space:
\begin{subequations}
\begin{gather}
  \mathcal{L}_\mathcal{U} n = 0
  , \label{L_U n} \\
  \mathcal{L}_\mathcal{U} \lambda^k =
  \mathcal{L}_\mathcal{U} \sigma_k  = 0
  , \label{L_U lambda,sigma} \\
  \mathcal{L}_\mathcal{U} \varphi = - i_\mathcal{U} \mathcal{P} + c^{-1} \theta
  , \label{L_U varphi} \\
  \dd \ast \dd A = \frac{e}{c} \, i_{\mathcal{U}} n
  .
\end{gather}
\end{subequations}

In 3-dimensional space, both $n$ and $\omega_1 \wedge \dd\sigma_2$ are 3-forms.
However, in 4-dimensional spacetime, $n$ is a 4-form and $\omega_1 \wedge \dd\sigma_2$ is a 3-form.
For non-relativistic physical quantities, the integration domain $\Omega(t)$ for integrating some density is contained in a hyperplane with some $x^0$ fixed ($t$-plane).
Using this, we can construct a non-relativistic generalized enstrophy $Q(t)$ in a relativistic setting, by performing a projection-like operation of taking the wedge product with $\dd x^0$:
\begin{align}
  Q(t)
  := \int_{\Omega(t)} f(\vartheta) \, i_{\partial_0} n
  \label{semi-Q def}
\end{align}
with a scalar $\vartheta$ and a 2-form $\omega_1$
\begin{align}
  \vartheta
  &:= \frac{ (\dd x^0 \wedge \omega_1 \wedge \dd\sigma_2)^\ast }{ (\dd x^0 \wedge i_{\partial_0} n)^\ast }
    = \frac{ (\dd x^0 \wedge \omega_1 \wedge \dd\sigma_2)^\ast }{ n^\ast }
  , \\
  \omega_1 &:= \dd\lambda^1 \wedge \dd\sigma_1
  .
\end{align}
In the following, we will call $Q(t)$ a \emph{semi-relativistic generalized enstrophy}
\footnote{
  The adjective ``semi-relativistic'' is in the spirit of ``the semi-relativistic helicity'' in Ref.~\onlinecite[Sec. III.A]{YoshidaKawazuraYokoyama2014}.
}.
In $\vartheta$, only the spatial component of $\omega_1 \wedge \dd\sigma_2$ is extracted by taking the wedge product with $\dd x^0$, so the semi-relativistic generalized enstrophy $Q(t)$ is consistent with the non-relativistic one $\widetilde{Q}(t)$.

If we calculate the time derivative of $Q(t)$ as in Appendix~\ref{sub: Time Evolution of the Semi-Relativistic Enstrophy}, we obtain the following result:
\begin{align}
  \dv{t} Q(t)
  &
  = \int_{\Omega(t)} \bigl\{
      f'(\vartheta) \,c\, (\dd\log\gamma \wedge \omega_1 \wedge \dd\sigma_2)^\ast
  \notag \\ & \qquad\qquad\qquad\qquad
      - f(\vartheta) \, n^\ast u(\log\gamma)
    \bigr\} \dd^3 x
  \label{dQdt} ,
\end{align}
where $u(\log\gamma)$ in the second term on the right-hand side is interpreted such that a vector field $u$ is differentiating $\log\gamma$, that is, $u(\log\gamma) = u^\mu \partial_\mu(\log\gamma)$.

The integrand of the left-hand side of equation (\ref{dQdt}) is proportional to the derivative of $\log\gamma$.
In the non-relativistic limit $\gamma \to 1$, $\dv{t} Q(t)$ goes to zero, but in general
$Q(t)$ is no longer a constant.
This is thought to be due to the correction of the volume area for integration by Lorentz contraction.
A similar phenomenon also occurs for helicity and is called the ``relativistic baroclinic effect'' in Ref.~\onlinecite{YoshidaKawazuraYokoyama2014}.
The fact that $Q(t)$ coincides with $\widetilde{Q}(t)$ in the non-relativistic limit is also parallel to the argument for helicity.

\section{Relativistic Generalized Enstrophy}
\label{sec: Relativistic Generalized Enstrophy}

\begin{figure}
  \begin{center}
    \begin{minipage}[t]{0.45\linewidth}
      \begin{center}
        \includegraphics[width=\linewidth]{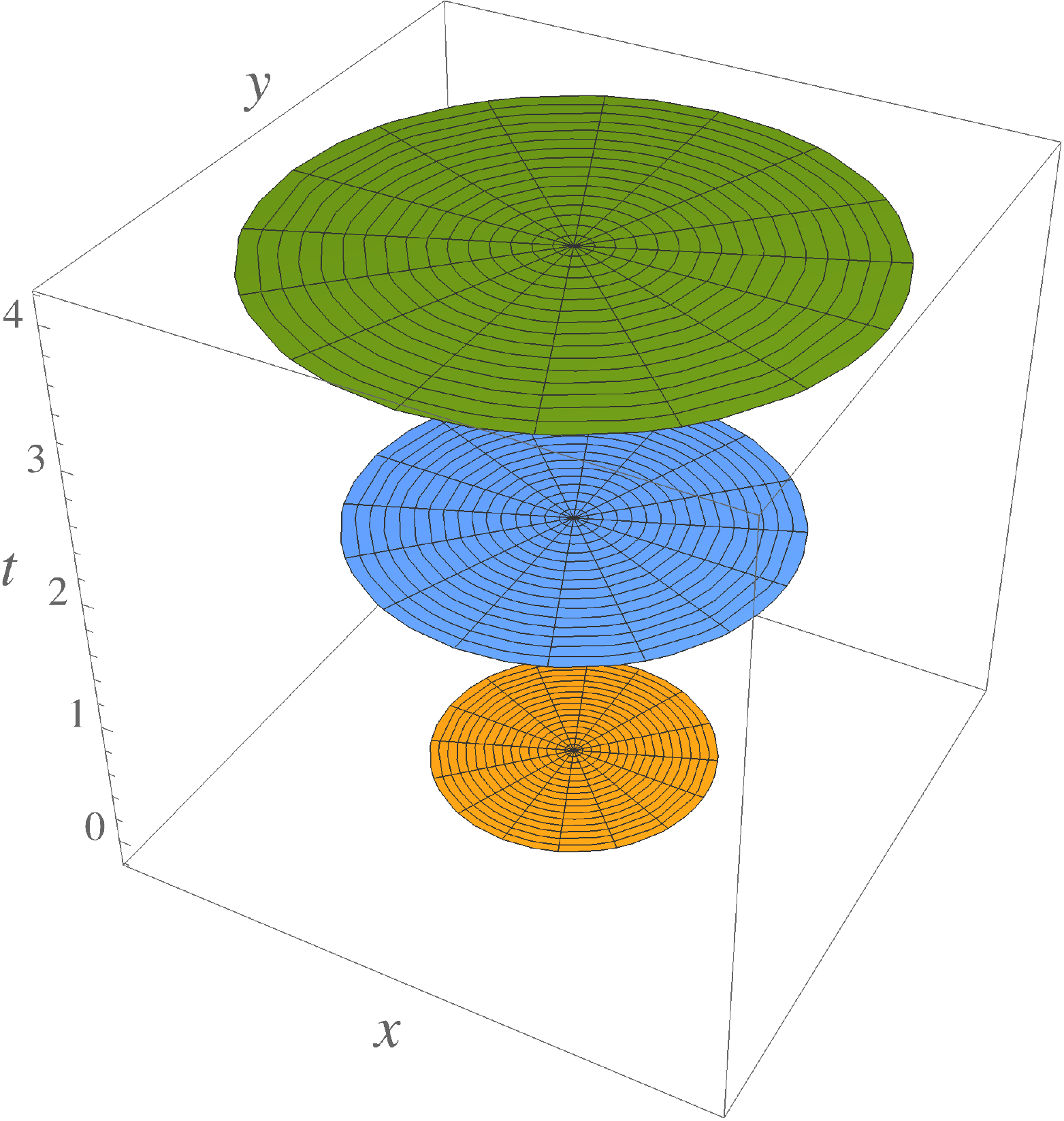}
      \end{center}
    \end{minipage}
    \begin{minipage}[t]{0.45\linewidth}
      \begin{center}
        \includegraphics[width=\linewidth]{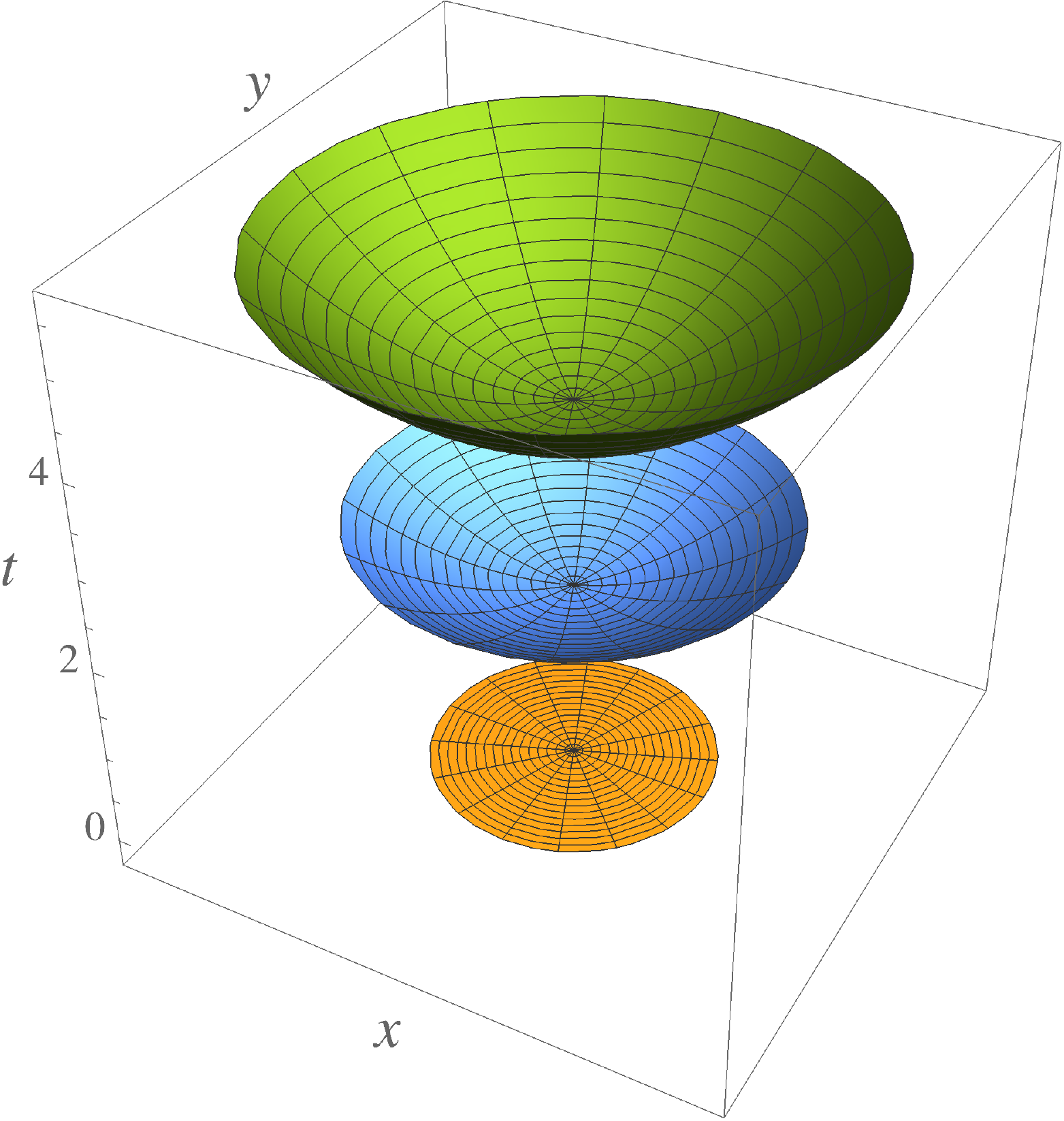}
      \end{center}
    \end{minipage}
  \end{center}
  \caption{
    A conceptual drawing of $t$-plane (left) and $s$-plane (right) of the fluid flowing outward radially from the origin.
    The $t$-plane is a hyperplane in space-time at constant time $t$, and used for the integral domain of the semi-relativistic enstrophy $Q(t)$.
    On the other hand, the $s$-plane is a curved hypersurface in space-time reflecting the intrinsic time $s$ of the fluid, and used for the integral domain of the relativistic enstrophy $\mathfrak{Q}(s)$.
    In this figure, the fluid farther from the origin is faster, so more time $t$ has passed on the outer side.
  }
  \label{t/splane.pdf}
\end{figure}

The semi-relativistic generalized enstrophy does not conserve because it is a physical quantity to be observed non-relativistically, even though we are considering a relativistically time-evolving field.
As in the case of helicity, we want to redefine generalized enstrophy to fit the relativistic picture.
We expect that the appropriate integral range $V(s)$ is a 3D hyper-surface in $M$ defined by $V(s) := \mathcal{T}_{\mathcal{U}}(s) V_0$, called an ``$s$-plane'' where $s$ is the proper time.
Figure \ref{t/splane.pdf} depicts the difference between the $t$-plane and $s$-plane.

In defining the generalized enstrophy density, we must define a scalar $\vartheta$ as an argument of the arbitrary function $f$.
In 3D space, it was easy to compare 3-form $n$ with 3-form $\omega_1 \wedge \dd\sigma_2$.
However, when we consider 4-dimensional space-time, the comparison is not obvious.
Now $n$ is a 4-form, but $\omega_1 \wedge \dd\sigma_2$ is still a 3-form.
We were unable to find a suitable way to compare $n$ with a 4-form made by taking the wedge product of $\omega_1 \wedge \dd\sigma_2$ and some 1-form.

We choose to compare the two 3-forms $\omega_1 \wedge \dd\sigma_2$ and $i_{\mathcal{U}} n$.
The linear space $\Omega^3(M)$ is 4 dimensional, and $\omega_1 \wedge \dd\sigma_2$ and $i_{\mathcal{U}} n$ are not necessarily parallel.
In order to compare them, we use a coordinate map on $V(s)$.
We hope that once the two 3-forms are pulled back, one is a scalar multiple of the other as 3-forms on a 3D submanifold.
A natural coordinate map on $V(s)$ seems to be the diffeomorphism $\mathcal{T}_{\mathcal{U}}(s) : V_0 \to V(s)$.

We must verify that the derivative by $s$ of the pullback by $\mathcal{T}_{\mathcal{U}}(s)$ and the Lie derivative $\mathcal{L}_{\mathcal{U}}$ are connected.
For a scalar function $g \in C^\infty(M)$,
\begin{align}
  \dv{s} \mathcal{T}_{\mathcal{U}}(s)^\ast g
  &
  = \dv{s} \, g \circ \mathcal{T}_{\mathcal{U}}(s)
  = \pdv{g}{x^\mu} \qty[ \dv{s} \mathcal{T}_{\mathcal{U}}(s) ]^\mu
  \notag \\ &
  = \pdv{g}{x^\mu} \mathcal{U}^\mu|_{\mathcal{T}_{\mathcal{U}}(s)}
  = \left. i_{\mathcal{U}} \dd g \right|_{\mathcal{T}_{\mathcal{U}}(s)}
  \notag \\ &
  = \left. \mathcal{L}_{\mathcal{U}} g \right|_{\mathcal{T}_{\mathcal{U}}(s)}
  = {\mathcal{T}_{\mathcal{U}}(s)}^\ast (\mathcal{L}_{\mathcal{U}} g)
  .\label{d/ds <-> LU func}
\end{align}
Any $k$-form can be written as a linear sum of exterior derivatives of scalar functions combined with wedge products.
Since a pullback is commutative with a wedge product, and both $\dv{s}$ and $\mathcal{L}_{\mathcal{U}}$ satisfy the Leibniz rule and are commutative with an exterior derivative, equation (\ref{d/ds <-> LU func}) holds for general $k$-forms (not only for 0-forms):
\begin{align}
  \dv{s} \mathcal{T}_{\mathcal{U}}(s)^\ast \omega
  = {\mathcal{T}_{\mathcal{U}}(s)}^\ast (\mathcal{L}_{\mathcal{U}} \omega)
  \label{d/ds <-> LU}
\end{align}
for any $\omega \in \Omega^k(M)$.
Based on this fact, the \emph{relativistic enstrophy} $\mathfrak{Q}(s)$ is defined as follows:
\begin{align}
  \mathfrak{Q}(s)
  := c^{-1} \int_{V_0} f(\vartheta) \, \mathcal{T}_{\mathcal{U}}(s)^\ast (i_{\mathcal{U}} n)
  \label{relativistic_enstrophy}
\end{align}
with a scalar function
\begin{align}
  \vartheta
  := c \,
      \frac{ \qty( \mathcal{T}_{\mathcal{U}}(s)^\ast (\omega_1 \wedge \dd\sigma_2) )^\ast }
            { \qty( \mathcal{T}_{\mathcal{U}}(s)^\ast ( i_{\mathcal{U}} n ) )^\ast }
  \label{relativistic_vartheta} .
\end{align}
Note that in the definition of $\vartheta$, $\mathcal{T}_{\mathcal{U}}(s)^\ast (\omega_1 \wedge \dd\sigma_2)$ and $\mathcal{T}_{\mathcal{U}}(s)^\ast ( i_{\mathcal{U}} n )$ are 3-forms on a 3-dimensional manifold $V_0 \subset \mathbb{R}^3$, which become 0-forms if we let Hodge star $\ast$ act on them.
Since
\begin{align}
  \mathcal{L}_{\mathcal{U}} (i_{\mathcal{U}} n)
  = \mathcal{L}_{\mathcal{U}} (\omega_1 \wedge \dd\sigma_2)
  = 0
\end{align}
can be derived from the time evolution of Clebsch parameters from equations (\ref{L_U n})--(\ref{L_U varphi}),
\begin{align}
  \dv{s} \mathcal{T}_{\mathcal{U}}(s)^\ast ( i_{\mathcal{U}} n )
  = \dv{s} \mathcal{T}_{\mathcal{U}}(s)^\ast ( \omega_1 \wedge \dd\sigma_2 )
  = 0
\end{align}
can be obtained from equation (\ref{d/ds <-> LU}).
Therefore, the conservation law of the relativistic enstrophy $\mathfrak{Q}(s)$ follows, i.e.
\begin{align}
  \dv{s} \mathfrak{Q}(s) = 0
  .
\end{align}

The definition of $\mathfrak{Q}(s)$ in equations (\ref{relativistic_enstrophy}), (\ref{relativistic_vartheta}) can be written symbolically as follows:
\begin{align}
  \mathfrak{Q}(s)
  &
  := c^{-1} \int_{V(s)} f(\vartheta) \, i_{\mathcal{U}} n
  , \\
  \vartheta
  &
  := c \,
      \frac{ \omega_1 \wedge \dd\sigma_2 }
            { i_{\mathcal{U}} n }
  .
\end{align}
This simple formalism would make it easier to understand the definition and to compare it with conventional enstrophy (\ref{conventional_enstrophy}).
However, this is only symbolic and ill-defined, and the actual calculation is as shown in equations (\ref{relativistic_enstrophy}), (\ref{relativistic_vartheta}).

In the following, 
we will check the correspondence of the semi-relativistic generalized enstrophy and the relativistic generalized enstrophy in the non-relativistic limit.
This is one of the reasons for the validity of the definition of the relativistic generalized enstrophy.
It is sufficient to verify that the semi-relativistic enstrophy is identical with the physical quantity where all $\mathcal{U}$ in the definition of the relativistic generalized enstrophy are changed to $u$.

First, an $s$-plane $V(s)$ becomes a $t$-plane $\Omega(t) := \mathcal{T}_{u}(t) \, \Omega_0$ where $V_0 = \Omega_0$, that is, the integral range co-moving with the fluid is flat at any time $t$ (or $s$).
Since it is necessary to distinguish between differential forms on 3-dimensional manifolds and on 4-dimensional manifolds when acting on the Hodge star, we make the inclusion map $\iota : \mathbb{R}^3 \supset \Omega(t) \to \Omega(t) \subset \mathbb{R}^4, \boldsymbol{x} \mapsto (ct, \boldsymbol{x})$ explicit in the following.
The diffeomorphism $\mathcal{T}_u(t)$ can be decomposed into $\iota$ and $\widetilde{\mathcal{T}}_u(t) : \mathbb{R}^3 \supset \Omega_0 \to \Omega(t) \subset \mathbb{R}^3$, i.e., $\mathcal{T}_u(t) = \iota \circ \widetilde{\mathcal{T}}_u(t)$.

We will see how the pullback by $\iota$ affects differential forms.
For a basis of 1-forms with the global coordinate $(x^0,x^1,x^2,x^3)$, the relations
$\iota^\ast \dd x^0 = 0$, $\iota^\ast \dd x^i = \dd x^i$ hold.
Any 3-form $\omega \in \Omega^3(M)$ can be written as
\begin{align}
  \omega
  &= a^0 \, \dd x^1 \!\wedge\! \dd x^2 \!\wedge\! \dd x^3
    + a^1 \, \dd x^0 \!\wedge\! \dd x^2 \!\wedge\! \dd x^3
  \notag \\ & \qquad
    + a^2 \, \dd x^0 \!\wedge\! \dd x^1 \!\wedge\! \dd x^3
    + a^3 \, \dd x^0 \!\wedge\! \dd x^1 \!\wedge\! \dd x^2
\end{align}
for some functions $a^0$, $a^1$, $a^2$, $a^3$.
When written down in this way, we see that the equation
\begin{align}
  (\iota^\ast \omega)^\ast = a^0 = \iota^\ast (\dd x^0 \wedge \omega)^\ast
  \label{Hodge of 3-forms}
\end{align}
holds because
\begin{align}
  \iota^\ast \omega &= a^0 \, \dd x^1 \!\wedge\! \dd x^2 \!\wedge\! \dd x^3
  , \\
  \dd x^0 \wedge \omega &= a^0 \, \dd x^0 \!\wedge\! \dd x^1 \!\wedge\! \dd x^2 \!\wedge\! \dd x^3
  .
\end{align}
Using equation (\ref{Hodge of 3-forms}), we can calculate the numerator and denominator of what corresponds to $\vartheta$.
\begin{align}
  \mathcal{T}_u(t)^\ast (\omega_1 \wedge \dd\sigma_2)
  &= \widetilde{\mathcal{T}}_u(t)^\ast \qty[ \, \iota^\ast (\omega_1 \wedge \dd\sigma_2) \, ]
  \notag \\
  &= \widetilde{\mathcal{T}}_u(t)^\ast \qty[ \qty( \iota^\ast (\omega_1 \wedge \dd\sigma_2) )^\ast \mathrm{vol}^3 ]
  \notag \\
  &= \widetilde{\mathcal{T}}_u(t)^\ast \qty[ \, \iota^\ast (\omega_1 \wedge \dd\sigma_2) \, ]^\ast
      \, \widetilde{\mathcal{T}}_u(t)^\ast \mathrm{vol}^3
  \notag \\
  &= \qty[ \widetilde{\mathcal{T}}_u(t)^\ast ( \dd x^0 \wedge \omega_1 \wedge \dd\sigma_2)^\ast ]
      \, \widetilde{\mathcal{T}}_u(t)^\ast \mathrm{vol}^3
  .
\end{align}
In the same way,
\begin{align}
  \mathcal{T}_u(t)^\ast i_u n
  = \qty[ \widetilde{\mathcal{T}}_u(t)^\ast ( \dd x^0 \wedge i_u n )^\ast ]
      \, \widetilde{\mathcal{T}}_u(t)^\ast \mathrm{vol}^3
  .
\end{align}
If we apply the Hodge star to both and take the ratio, the terms resulting from $\widetilde{\mathcal{T}}_u(t)^\ast \mathrm{vol}^3$ (corresponding to the Jacobian) cancel each other out:
\begin{align}
  \mathcal{T}_u(t)^\ast
  \frac{ (\dd x^0 \wedge \omega_1 \wedge \dd\sigma_2)^\ast }
        { (\dd x^0 \wedge i_u n)^\ast }
  = \frac{ \qty( \mathcal{T}_u(t)^\ast (\omega_1 \wedge \dd\sigma_2) )^\ast }
          { \qty( \mathcal{T}_u(t)^\ast i_u n )^\ast }
  .
\end{align}
From the above, we find the correspondence between the semi-relativistic enstrophy $Q(t)$ and the relativistic generalized enstrophy $\mathfrak{Q}(s)$.

Since $\mathfrak{Q}(s)$ is relativistically conserved and consistent with $Q(t)$ in the non-relativistic limit, we can conclude that we have defined a relativistically valid Casimir $\mathfrak{Q}(s)$ equivalent to the enstrophy.
The fact that a Casimir is present not only in non-relativistic plasmas but also in relativistic plasmas suggests that the complexity of plasma behavior does not change significantly between non-relativistic and relativistic plasmas.

As discussed in Ref.~\onlinecite[Sec. 5]{YoshidaMorrison2017_springer}, a domain with $\lambda^2 = 0$ can be viewed as a quasi-particle.
In particular, when $f = \mathrm{id}_{\mathbb{R}}$, $\mathfrak{Q}(s)$ is interpreted as the number of vortex filaments penetrating a $\sigma_2$ constant surface on the hypersurface $V(s)$.

\section{Conclusion}
\label{sec: Conclusion}

There are three main results in this paper.
First, we have formulated the special relativistic Hamiltonian mechanics of a barotropic charged fluid by invoking the Clebsch parametrization of the field variables;
the Poisson manifold (phase space) is the Hilbert space of the Clebsch parameters, in which the charged fluid system is embedded as a subalgebra through the Clebsch reduction.
The Clebsch parameters are useful to specify topological constraints on the fluid variables.

Second, we have evaluated the time evolution of the conventional generalized enstrophy.
However, this is not a proper relativistic quantity, so its non-conservation demonstrates the relativistic effect on the vorticity observed in the reference frame.
We find that the usual Kelvin's circulation theorem does not hold in a relativistic charged fluid.

Third, we have derived a relativistically conserving generalized enstrophy.
We are interested in the ratio of two 3-forms $\omega_1 \wedge \dd\sigma_2$ and $i_{\mathcal{U}} n$ (not necessarily parallel) in the 4-dimensional time-space.
To solve this problem, we take the ratio by projecting it in the direction orthogonal to the hypersurface $V(s)$, using the pullback by the diffeomorphism $\mathcal{T}_{\mathcal{U}}(s)$.
The two 3-forms can be thought of as analogs of four-currents, and the operation of projecting orthogonally to the hypersurface $V(s)$ corresponds to extracting only the charge density (the 0th component of the four-current) of the two on $V(s)$ and comparing them.

The relativistic effect causes the loss of simultaneity allowing the elapsed proper time $s$ of fluid elements depend on the location.
The relativistically conservative enstrophy must be evaluated for the region in the $s$-plane so that the elapsed proper time $s$ of each fluid element is constant.

\section*{Acknowledgments}
\label{sec: Acknowledgments}

This work was supported by JSPS KAKENHI grant number 17H01177.  The authors acknowledge the useful suggestion of Dr. Naoki Sato.

\section*{Conflict of Interest Disclosure}
\label{sec: Conflict of Interest Disclosure}

The authors have no conflicts to disclose.

\section*{Data Availability Statement}
\label{sec: Data Availability Statement}

Data sharing is not applicable to this article as no new data were created or analyzed in this study.

\appendix

\section{Notation Comparison of Differential Geometry and Tensor Analysis}
\label{sec: Notation Comparison of Differential Geometry and Tensor Analysis}

This section provides an overview of the differential geometry notation used in this paper.
Most of them are standard notations in differential geometry, but are briefly summarized to facilitate review of their meanings and properties.
Some important formulas written in differential geometry notation will be shown together with formulas in tensor analysis notations (or vector calculus notations) in Table~\ref{tab:Notation comparison in non-relativistic description.} and \ref{tab:Notation comparison in relativistic description.}.
Note that $\epsilon^{\mu\nu\rho\sigma}$ in Table~\ref{tab:Notation comparison in relativistic description.} represents the Levi-Civita symbol in four dimensions, which corresponds to the cross product in Table~\ref{tab:Notation comparison in non-relativistic description.}.

\begin{table}
  \caption{Notation comparison in non-relativistic description.}
  \label{tab:Notation comparison in non-relativistic description.}
  \begin{ruledtabular}
    \begin{tabular}{cc}
      Differential geometry notations  &   Vector calculus notations  \\ \hline
      $i_{\boldsymbol{V}} \mathcal{P}$  &   $\boldsymbol{V} \vdot \boldsymbol{P}$
      \\
      $\mathcal{L}_{\boldsymbol{V}} n$, \quad $\mathcal{L}_{\boldsymbol{V}} \Lambda^k$  &   $\nabla \vdot (\rho \boldsymbol{V})$, \quad $\nabla \vdot (\Lambda^k \boldsymbol{V})$
      \\
      $\mathcal{L}_{\boldsymbol{V}} \varphi$, \quad $\mathcal{L}_{\boldsymbol{V}} \sigma_k$, \quad $\mathcal{L}_{\boldsymbol{V}} \lambda^k$  & $\boldsymbol{V} \vdot \nabla \varphi$, \quad $\boldsymbol{V} \vdot \nabla \sigma_k$, \quad $\boldsymbol{V} \vdot \nabla \lambda^k$
      \\
      $\omega_1 := \dd\lambda^1 \!\wedge \dd\sigma_1$  &   $\boldsymbol{\omega}_1 := \nabla\lambda^1 \!\times\! \nabla\sigma_1$
      \\
      $\displaystyle \vartheta_1 := \frac{ \qty( \omega_1 \wedge \dd\sigma_2 )^\ast }{n^\ast}$  &   $\displaystyle \vartheta_1 := \frac{ \boldsymbol{\omega}_1 \vdot \nabla\sigma_2 }{\rho} = \frac{ \nabla\lambda^1 \!\times\! \nabla\sigma_1 \vdot \nabla\sigma_2 }{\rho}$
      \\
    \end{tabular}
  \end{ruledtabular}
\end{table}

\begin{table}
  \caption{Notation comparison in relativistic description.}
  \label{tab:Notation comparison in relativistic description.}
  \begin{ruledtabular}
    \begin{tabular}{cc}
      Differential geometry notations  &   Tensor analysis notations  \\ \hline
      $\dd\mathcal{P}$  &   $\partial_\mu \mathcal{P}_\nu - \partial_\nu \mathcal{P}_\mu$
      \\
      $\mathcal{K} := \mathcal{P} \wedge \dd\mathcal{P}$  &   $\mathcal{K}^\mu := \epsilon^{\mu\nu\rho\sigma} \mathcal{P}_\nu \partial_\rho \mathcal{P}_\sigma$
      \\
      $\mathcal{L}_\mathcal{U} n$ &    $\partial_\mu (n\,\mathcal{U}^\mu)$
      \\
      $\mathcal{L}_\mathcal{U} \varphi$, \quad $\mathcal{L}_\mathcal{U} \lambda^k$, \quad $\mathcal{L}_\mathcal{U} \sigma_k$ &   $\mathcal{U}^\mu \partial_\mu \varphi$, \quad $\mathcal{U}^\mu \partial_\mu \lambda^k$, \quad $\mathcal{U}^\mu \partial_\mu \sigma_k$
      \\
      $i_\mathcal{U} \mathcal{P}$ &   $\mathcal{U}^\nu \mathcal{P}_\nu$
      \\
      $\mathcal{L}_\mathcal{U} \mathcal{P}$ &   $\mathcal{U}^\nu \partial_\nu \mathcal{P}^\mu + \mathcal{P}^\nu \partial_\mu \mathcal{U}_\nu$
      \\
      $i_{\mathcal{U}} n$ &   $n\,\mathcal{U}^\mu$
      \\
      $\omega_1 := \dd\lambda^1 \wedge \dd\sigma_1$  &   $(\omega_1)_{\mu\nu} := \partial_\mu \lambda^1 \partial_\nu \sigma_1 - \partial_\nu \lambda^1 \partial_\mu \sigma_1$
      \\
      $\dd x^0 \wedge \omega_1 \wedge \dd\sigma_2$  &   $\epsilon^{0\nu\rho\sigma} (\omega_1)_{\nu\rho} \partial_\sigma \sigma_2$
      \\
      $\dd \ast \dd A = \frac{e}{c} \, i_{\mathcal{U}} n$ &   $\Box A^\mu - \partial^\mu (\partial_\nu A^\nu)
      = \frac{e}{c} n \,\mathcal{U}_\mu$
    \end{tabular}
  \end{ruledtabular}
\end{table}

Let $M$ be a differential manifold. 
A (smooth) \emph{differential $k$-form} is an alternating $k$-linear map from a $k$-tuple of (smooth) vector fields to a (smooth) scalar function. 
Denote the set of all smooth scalar functions by $C^\infty(M)$, the set of all smooth vector fields by $\mathfrak{X}(M)$, and the set of all smooth $k$-form by $\Omega^k(M)$. 
A 1-form is integrated over curves, a 2-form over surfaces, and a 3-form over volumes. 
The velocity field of a fluid is expressed as a vector field, but most other physical quantities are expressed as $k$-forms.
For example, for non-relativistic theory, a (canonical) momentum field and an EM vector potential are 1-forms, a vorticity and a magnetic field are 2-forms, and a mass density is a 3-form.

The \emph{wedge product} $\wedge$ is a bilinear map that creates a $(k+l)$-form $\omega \wedge \tau$ from a $k$-form $\omega$ and an $l$-form $\tau$.
This operation satisfies anticommutativity in the sense that $\omega \wedge \tau = (-1)^{kl} \tau \wedge \omega$.
In 3-dimensional space, the wedge product between 1-forms corresponds to the cross product $\boldsymbol{A} \times \boldsymbol{B}$, and the wedge product between a 1-form and a 2-form corresponds to the dot product $\boldsymbol{A} \vdot \boldsymbol{B}$ in vector calculus.

The \emph{exterior derivative} $\dd$ is a linear map from $k$-forms to $(k+1)$-forms satisfying a kind of Leibniz rule:
$\dd(\omega \wedge \tau) = (\dd\omega) \wedge \tau + (-1)^k \omega \wedge \dd\tau$ for $\omega \in \Omega^k(M)$, $\tau \in \Omega^l(M)$.
In three dimensions, it corresponds to grad, rot, and div in vector calculus, i.e.,
\[
  \begin{CD}
    \Omega^0(M) @>{\dd}>> \Omega^1(M) @>{\dd}>> \Omega^2(M) @>{\dd}>> \Omega^3(M) \\
  @V{\simeq}VV    @V{\simeq}VV    @V{\simeq}VV    @V{\simeq}VV \\
    C^\infty(M) @>{\mathrm{grad}}>> \mathfrak{X}(M) @>{\mathrm{rot}}>> \mathfrak{X}(M) @>{\mathrm{div}}>> C^\infty(M)
  .
  \end{CD}
\]
The property $\dd \circ \dd = 0$ are valid, which corresponds to $\mathrm{rot}\,\mathrm{grad} = 0$ and $\mathrm{div}\,\mathrm{rot} = 0$.

The \emph{interior multiplication} denoted by $i_V \omega$ is defined to be the contraction of a differential form $\omega \in \Omega^k(M)$ with a vector field $V \in \mathfrak{X}(M)$.
For $k \geq 1$, the $(k-1)$-form $i_V \omega$ is defined by
\begin{align}
  (i_V \omega)(X_2, \cdots , X_k) := \omega(V, X_2, \cdots , X_k)
\end{align}
for $X_2, \cdots , X_k \in \mathfrak{X}(M)$,
and $i_V \omega := 0$ for $\omega \in \Omega^0(M)$.
For a 1-form $\alpha \in \Omega^1(M)$, $i_V \alpha \in \Omega^0(M)$ is a dot product of $V \in \mathfrak{X}(M)$ and a vector corresponding to $\alpha$,
although the operation in tensor analysis notation is a little more complicated when $k \geq 2$.

The \emph{Lie derivative} $\mathcal{L}_X$ (of a differential form) is a linear map from $k$-forms to $k$-forms satisfying the Leibniz rule $\mathcal{L}_X(\omega\wedge\tau) = (\mathcal{L}_X \omega) \wedge \tau + \omega \wedge (\mathcal{L}_X \tau)$ for $\omega \in \Omega^k(M)$, $\tau \in \Omega^l(M)$.
$\mathcal{L}_X \omega$ evaluates the change of the differential form $\omega$ along the vector field $X$.
The Lie derivative for 0-forms $\mathcal{L}_V \varphi$ matches the convection term $\boldsymbol{V}\vdot\nabla \varphi$ of the material derivative.
The Lie derivative of a differential form relates the exterior derivative and the interior multiplication by the Cartan formula $\mathcal{L}_X = \dd i_X + i_X \dd$. 

Let $M$ and $N$ be differential manifolds, and $F: N \to M$ be a smooth map. 
For a $k$-form $\omega \in \Omega^k(M)$ on $M$, when $F$ is regarded as a coordinate transformation, the $k$-form on $N$ corresponding to $\omega$ by $F$ is called the \emph{pullback} and written as $F^\ast \omega$.
For a 0-form $h \in C^\infty(M)$, $F^\ast h = h \circ F$.
The pullback map is a linear operator which commutes with the exterior derivative $F^\ast (\dd\omega) = \dd (F^\ast \omega)$ and which is distributive for wedge product $F^\ast (\omega\wedge\tau) = (F^\ast \omega) \wedge (F^\ast \tau)$. 

All operations introduced in this subsection are \emph{intrinsic}, i.e., coordinate invariant.
For more precise definitions and detailed properties, see, for example, Tu~\cite{Tu2011manifold}
(at least the differential geometry notations in this paper is the same as in Tu~\cite{Tu2011manifold}).

\section{Intermediate Calculations}
\label{sec: Intermediate Calculations}

For the convenience of the reader, we describe details of calculations omitted in the main text to avoid tedious steps of derivations.

\subsection{Variation of the Action Functional to Derive Clebsch Parametrization}
\label{sub: Variation of the Action Functional to Derive Clebsch Parametrization}

We start by calculating the variation of the action functional defined in equation (\ref{action}),
\begin{align}
  \delta S
  &= \int \delta n \qty[ \frac{L_F}{n} - n \,\mathcal{E}'(n) \, \frac{\mathcal{U}_\mu \,\mathcal{U}^\mu}{
    c^2} ] \dd^4 x
  \notag \\ & \quad
    - \int \delta \,\mathcal{U}^\mu \qty[
        \frac{e}{c} A_\mu
        + \partial_\mu \varphi
        + \sum_k \lambda^k \partial_\mu \sigma_k
      ] n \, \dd^4 x
  \notag \\ & \quad
    - \int n \,\mathcal{U}^\mu \partial_\mu \delta\varphi \, \dd^4 x
    - \sum_k \int \delta\lambda^k \!\cdot\! n \,\mathcal{U}^\mu \partial_\mu \sigma_k \,\dd^4 x
  \notag \\ & \quad
    - \sum_k \int n \lambda^k \mathcal{U}^\mu \partial_\mu \delta\sigma_k \,\dd^4 x
    - \int \delta A^\mu \!\cdot\! \frac{e}{c} n \,\mathcal{U}_\mu \,\dd^4 x
  \notag \\ & \quad
    + \int \delta \qty( -\frac{1}{4} F_{\mu\nu} F^{\mu\nu} ) \dd^4 x
    + \mathcal{O}(\delta^2)
  .
  \label{delta S}
\end{align}

Because the action functional is expressed in terms of $\mathcal{U}^\mu$ without using $v^j$, we will first examine how the change in $v^j$ affects $\mathcal{U}^\mu$.
Recalling that $\mathcal{U}^\mu$ and $v^j$ are linked together in equations (\ref{mathcalU})--(\ref{gamma}), we have
\begin{align}
  \delta \,\mathcal{U}^0 &= \frac{\gamma^3}{c} v^j \delta v^j
  , &
  \delta \,\mathcal{U}^i &= \gamma \delta v^i + \frac{\gamma^3}{c^2} v^i v^j \delta v^j
  .
\end{align}
Here, we introduce a vector $W$ to avoid a long formula
\begin{align}
  W_\mu
  := \frac{e}{c} A_\mu
      + \partial_\mu \varphi
      + \sum_k \lambda^k \partial_\mu \sigma_k
  . \label{W}
\end{align}
With this notation, the second term of equation (\ref{delta S}) is
\begin{align}
  &
  - \int \delta \,\mathcal{U}^\mu \, n W_\mu \,\dd^4 x
  \notag \\
  &= - \int \delta v^i \!\cdot\! \frac{\gamma^2}{c^2} v^i \,n \,\mathcal{U}^\mu W_\mu \,\dd^4 x
      - \int \delta v^i \!\cdot\! \gamma \, n W_i \,\dd^4 x
  .
\end{align}

Furthermore, the last term of equation (\ref{delta S}) can be modified as follows:
\begin{align}
  \int \delta \qty( -\frac{1}{4} F_{\mu\nu} F^{\mu\nu} ) \,\dd^4 x
  &= \int \delta A_\mu \, \qty( \Box A^\mu - \partial^\mu (\partial_\nu A^\nu)) \,\dd^4 x
  .
\end{align}

To summarize the above, we obtain the Euler-Lagrange equations:
\begin{subequations}
\begin{align}
  & \text{by $\delta n$,} &&
  \frac{L_F}{n} = n \,\mathcal{E}'(n)
  \label{delta n (pre)}
  \\ & \text{by $\delta v^i$,} &&
  \qty(
    \frac{e}{c} A_i
    + \partial_i \varphi
    + \sum_k \lambda^k \partial_i \sigma_k
  )
  \notag \\ &&&
  - \frac{1}{c^2} \,\mathcal{U}_i \,\mathcal{U}^\mu
    \qty(
      \frac{e}{c} A_\mu
      + \partial_\mu \varphi
      + \sum_k \lambda^k \partial_\mu \sigma_k
    )
  = 0
  \label{delta vi (pre1)}
  \\ & \text{by $\delta \varphi$,} &&
  \partial_\mu\qty( n \,\mathcal{U}^\mu ) = 0
  \label{delta varphi (pre)}
  \\ & \text{by $\delta \lambda^k$,} &&
  \mathcal{U}^\mu \partial_\mu \sigma_k = 0
  \\ & \text{by $\delta \sigma_k$,} &&
  \partial_\mu \qty( n \lambda^k \mathcal{U}^\mu ) = 0
  , \qquad \therefore \,
  \mathcal{U}^\mu \partial_\mu \lambda_k = 0
  \\ & \text{by $\delta A_\mu$,} &&
  \Box A^\mu - \partial^\mu (\partial_\nu A^\nu)
  - \frac{e}{c} n \,\mathcal{U}_\mu = 0
  \label{delta A (pre)} .
\end{align}
\end{subequations}
Equations (\ref{delta varphi (pre)})--(\ref{delta A (pre)}) are nothing less than equations (\ref{delta varphi}), (\ref{delta lambda sigma}), (\ref{delta A}).
Using the relation
\begin{align}
  h(n)
  = \pdv{(n\mathcal{E}(n))}{n}
  = \mathcal{E}(n) + n \mathcal{E}'(n)
  ,
\end{align}
equation (\ref{delta n (pre)}) is equivalent to (\ref{delta n}).

We can organize equation (\ref{delta vi (pre1)}) further.
With the notation from equation (\ref{W}), equation (\ref{delta vi (pre1)}) can be rewritten as
\begin{align}
  W_i - \frac{1}{c^2} \,\mathcal{U}_i \,\mathcal{U}^\mu W_\mu = 0
  \label{delta vi (pre2)}
\end{align}
for $i = 1,2,3$.
By contracting both sides of equation (\ref{delta vi (pre2)}) with $\mathcal{U}^i$,
\begin{align}
  0
  &= \,\mathcal{U}^i W_i
    - \frac{1}{c^2} \,\mathcal{U}^i \,\mathcal{U}_i \,\mathcal{U}^\mu W_\mu
  \notag \\
  &= \,\mathcal{U}^\mu W_\mu - \,\mathcal{U}^0 W_0
    - \frac{1}{c^2} \qty( c^2 - \,\mathcal{U}_0 \,\mathcal{U}^0 ) \,\mathcal{U}^\mu W_\mu
  \notag \\
  &= - \,\mathcal{U}^0 \qty(
      W_0 - \frac{1}{c^2} \,\mathcal{U}_0 \,\mathcal{U}^\mu W_\mu
    )
  .
\end{align}
Because $\,\mathcal{U}^0 = c \gamma$ is positive at each point, equation (\ref{delta vi (pre2)}) is found to be valid even if we set $i=0$, i.e.
\begin{align}
  W_\nu - \frac{1}{c^2} \,\mathcal{U}_\nu \,\mathcal{U}^\mu W_\mu = 0
  \label{delta vi (pre3)}
\end{align}
for $\nu = 0,1,2,3$.

Now we introduce a matrix $q$ as follows:
\begin{align}
  q_\nu{}^\mu := \delta_\nu^\mu - \frac{1}{c^2} \,\mathcal{U}_\nu \,\mathcal{U}^\mu
\end{align}
with Kronecker delta $\delta_\nu^\mu$.
It follows that $q$ is a projection operator with $\mathrm{Ker} (q) = \mathrm{Span} \qty{ \,\mathcal{U}_\mu }$.
Since equation (\ref{delta vi (pre3)}) means $q W = 0$, it follows that $W_\mu + f \,\mathcal{U}_\mu = 0$ for some scalar function $f(x)$.
By contracting both sides of equation $W_\mu + f \,\mathcal{U}_\mu = 0$ with $\mathcal{U}^\mu$,
\begin{align}
  f c^2
  &
  = - \,\mathcal{U}^\mu \qty(
        \frac{e}{c} A_\mu
        + \partial_\mu \varphi
        + \sum_k \lambda^k \partial_\mu \sigma_k
    )
  \notag \\ &
  = \frac{L_F}{n} + \mathcal{E}(n)
  = \pdv{\qty( n \mathcal{E}(n) )}{n}
  .
\end{align}
Recalling the definition of molar enthalpy, we find that $f = h(n)/c^2$.
From the above, equation (\ref{delta v}) is derived from equation (\ref{delta vi (pre1)}).

\subsection{Derive the Generalized Ohm's Law from Clebsch Parametrization}
\label{sub: Derive the Generalized Ohm's Law from Clebsch Parametrization}

In this subsection, we derive equations (\ref{generalized_Ohm}) and (\ref{relativistic_Euler}) from equations (\ref{delta varphi})--(\ref{delta v}).
The left-hand side of equation (\ref{generalized_Ohm}) is
\begin{align}
  \frac{1}{c^2}
  \partial_\nu
  \qty[
    \, n h(n) \, \mathcal{U}^\mu \mathcal{U}^\nu 
  ]
  = - n \,\mathcal{U}^\nu \partial_\nu
    \qty(
      \frac{e}{c} A^\mu + \partial^\mu \varphi + \sum_k \lambda^k \partial^\mu \sigma_k
    )
  \label{first nhUU}
\end{align}
using equation (\ref{delta v}).
The two last terms can be transformed respectively as follows:
\begin{align}
  n \,\mathcal{U}^\nu \partial^\mu \partial_\nu \varphi
  = - n \, \partial^\mu \qty{ h(n) + \frac{e}{c} \,\mathcal{U}^\nu A_\nu }
    - n \, ( \partial^\mu \,\mathcal{U}^\nu ) \partial_\nu \varphi 
\end{align}
from equation (\ref{delta n}), and
\begin{align}
  n \,\mathcal{U}^\nu \partial_\nu \qty( \lambda^k \partial^\mu \sigma_k )
  = - n \lambda^k ( \partial^\mu \,\mathcal{U}^\nu ) \partial_\nu \sigma_k
\end{align}
from equation (\ref{delta lambda sigma}).
Applying these, we obtain
\begin{align}
  &
  \frac{1}{c^2} \partial_\nu
  \qty[
    \, n h(n) \, \mathcal{U}^\mu \mathcal{U}^\nu 
  ]
  \notag \\  & \,\,\,
  = \frac{e}{c} n \,\mathcal{U}_\nu \qty(
      \partial^\mu A^\nu - \partial^\nu A^\mu
    ) 
    + n \, \partial^\mu h(n)
    - \frac{1}{c^2} n h(n) \,\mathcal{U}_\nu ( \partial^\mu \,\mathcal{U}^\nu )
  .
\end{align}
Here the last term vanishes because
\begin{align}
  \mathcal{U}_\nu ( \partial^\mu \,\mathcal{U}^\nu )
  = \frac{1}{2} \partial^\mu 
    \qty( \,\mathcal{U}_\nu \,\mathcal{U}^\nu )
  = 0
\end{align}
Since the pressure $p$ and enthalpy $h$ have the relationship
\begin{align}
  \partial^\mu p
  &
  = \partial^\mu \qty{ n h(n) - n \mathcal{E}(n) }
  \notag \\ &
  = \partial^\mu \qty{ 
      n \pdv{(n\mathcal{E}(n))}{n}
    }
    - ( \partial^\mu n ) \pdv{( n \mathcal{E}(n) )}{n}
  \notag \\ &
  = n \partial^\mu \qty{ 
      \pdv{(n\mathcal{E}(n))}{n}
    }
  \notag \\ &
  = n \, \partial^\mu h(n)
  ,
\end{align}
the proof of equation (\ref{generalized_Ohm}) is now complete.
With a slight modification of equation  (\ref{generalized_Ohm}), equation (\ref{relativistic_Euler}) can also be obtained:
\begin{align}
  n \, \mathcal{U}^\nu \partial_\nu \mathcal{P}^\mu
  &
  = \frac{1}{c^2} \partial_\nu
    \qty[
      \, n h(n) \, \mathcal{U}^\mu \mathcal{U}^\nu 
    ]
    + \frac{e}{c} n \, \mathcal{U}^\nu \partial_\nu A^\mu
  \notag \\ &  
  = \frac{e}{c} n \, \mathcal{U}_\nu \partial^\mu A^\nu + \partial^\mu p
  .
\end{align}

\subsection{Time Evolution of the Semi-Relativistic Enstrophy}
\label{sub: Time Evolution of the Semi-Relativistic Enstrophy}

Considering that the integration domain $\Omega(t)$ is perpendicular to the $x^0$ direction, $Q(t)$ defined in equation (\ref{semi-Q def}) can be rewritten as
\begin{align}
  Q(t) = c^{-1} \int_{\Omega(t)} f(\vartheta) \, i_u n
  .
\end{align}

The non-relativistic plasma motion is represented by a diffeomorphism $\mathcal{T}_u(t)$ that is generated by the non-relativistic four-velocity $u$, i.e.,
\begin{align}
  \dv{t} \mathcal{T}_u(t) = u
  .
\end{align}
This is contrasted with the relativistic plasma motion being represented by a diffeomorphism $\mathcal{T}_{\mathcal{U}}(s)$.
Using the fact that the integration domain $\Omega(t)$ satisfies $\mathcal{T}_u(t) \Omega(0)$, we have
\begin{align}
  \dv{t} Q(t)
  &= c^{-1} \int_{\Omega(t)} f'(\vartheta) \qty(\mathcal{L}_u \vartheta) \, i_u n
    + c^{-1} \int_{\Omega(t)} f(\vartheta) \, \mathcal{L}_u \qty(i_u n)
  .
\end{align}

Let us calculate $\mathcal{L}_u \vartheta$ first.
Since we want to use the conservation law of Clebsch parameters from (\ref{L_U n}) and (\ref{L_U lambda,sigma}), we will consider the Lie derivative by $\mathcal{U}$ and then return to the Lie derivative by $u$.
About the numerator of $\vartheta$, 
\begin{gather}
  i_{\mathcal{U}} \dd x^0
  = \mathcal{U}^0
  = c \gamma
  , \\ \therefore
  \mathcal{L}_{\mathcal{U}} (\dd x^0 \wedge \omega_1 \wedge \dd\sigma_2)
  = c \, \dd\gamma \wedge \omega_1 \wedge \dd\sigma_2
  .
\end{gather}
In general, applying the Lie derivative $\mathcal{L}_X$ to both sides of a $n$-form $\alpha = \alpha^\ast \mathrm{vol}^n \in \Omega^n(M)$, we obtain 
\begin{align}
  (\mathcal{L}_X \alpha)^\ast &= \mathcal{L}_X (\alpha^\ast) + \alpha^\ast (\mathrm{div} X)
  ,
\end{align}
where the divergence $\mathrm{div} X$ of $X \in \mathfrak{X}(M)$ is defined as a scalar function which satisfies $\mathcal{L}_X \mathrm{vol}^n = (\mathrm{div} X) \, \mathrm{vol}^n$.
Applying this formula and the Leibniz rule to $\mathcal{L}_{\mathcal{U}}$, we obtain
\begin{align}
  \mathcal{L}_{\mathcal{U}} \vartheta
  &= \frac{ \mathcal{L}_{\mathcal{U}} (\dd x^0 \wedge \omega_1 \wedge \dd\sigma_2)^\ast }{ n^\ast }
      - \frac{ (\dd x^0 \wedge \omega_1 \wedge \dd\sigma_2)^\ast \mathcal{L}_{\mathcal{U}} (n^\ast) }{ (n^\ast)^2 }
  \notag \\
  &= \frac{ (c \, \dd\gamma \wedge \omega_1 \wedge \dd\sigma_2)^\ast }{ n^\ast }
  . \label{L_U vartheta}
\end{align}

Here we introduce a generally-applicable formula in order to connect $\mathcal{L}_{\mathcal{U}}$ and $\mathcal{L}_u$:
\begin{align}
  \mathcal{L}_{fX} \omega
  &= \dd f \wedge i_X \omega + f \mathcal{L}_{X} \omega
  \label{L_fX}
\end{align}
for any $f \in C^\infty(M)$, $X \in \mathfrak{X}(M)$, and $\omega \in \Omega^k(M)$ on a general $C^\infty$ manifold $M$.
By combining equations (\ref{L_U vartheta}) and (\ref{L_fX}), we get
\begin{align}
  \mathcal{L}_u \vartheta
  &= \gamma^{-1} \mathcal{L}_{ \mathcal{U} } \vartheta
  \notag \\
  &= \gamma^{-1} \frac{ (c \, \dd\gamma \wedge \omega_1 \wedge \dd\sigma_2)^\ast }{ n^\ast }
  \notag \\
  &= \frac{ (c \, \dd(\log\gamma) \wedge \omega_1 \wedge \dd\sigma_2)^\ast }{ n^\ast }
  .
\end{align}

Let us move on to the transformation of $\mathcal{L}_u \qty(i_u n)$.
Using equation (\ref{L_fX}), the particle number conservation law (\ref{L_U n}) can be modified as follows:
\begin{align}
  \mathcal{L}_u n
  &= \dd \qty(\gamma^{-1}) \wedge i_{ \mathcal{U} } n
  \notag \\
  &= \dd \qty(\gamma^{-1}) \wedge \qty( \gamma \, i_u n )
  \notag \\
  &= - \dd \qty(\log\gamma) \wedge i_u n
  .
\end{align}
Using the fact that $\mathcal{L}_X$ and $i_X$ are commutative in general,
\begin{align}
  \mathcal{L}_u \qty(i_u n)
  &= i_u \qty(\mathcal{L}_u n)
  \notag \\
  &= - i_u \qty( \dd \qty(\log\gamma) \wedge i_u n )
  \notag \\
  &= - u \qty(\log\gamma) \, i_u n
  .
\end{align}

From the above, we obtain the expression of the time evolution of the semi-relativistic enstrophy 
\begin{align}
  \dv{t} Q(t)
  &=  c^{-1} \int_{\Omega(t)} \biggl\{
        f'(\vartheta) \frac{ (c \, \dd(\log\gamma) \wedge \omega_1 \wedge \dd\sigma_2)^\ast }{ n^\ast }
  \notag \\ & \qquad\qquad\qquad\qquad\qquad\qquad
        - f(\vartheta) \, u \qty(\log\gamma)
      \biggr\} \, i_u n
  \notag \\
  &=  \int_{\Omega(t)} \bigl\{
        f'(\vartheta) \,c\, (\dd\log\gamma \wedge \omega_1 \wedge \dd\sigma_2)^\ast
  \notag \\ & \qquad\qquad\qquad\qquad
      - f(\vartheta) \, n^\ast u(\log\gamma)
      \bigr\} \,\dd^3 x
  .
\end{align}

\bibliography{reference_papers.bib}

\end{document}